\documentclass[twoside,11pt]{article}

\usepackage[abbrvbib, preprint]{jmlr2e}

\usepackage{jmlr2e}
\usepackage{booktabs} 
\usepackage{multirow} 
\usepackage{amsmath} 
\usepackage{adjustbox}
\usepackage{xcolor}
\usepackage{array}
\usepackage{authblk} 

\definecolor{wordcolor1}{rgb}{0.796,0.627,0.537}
\definecolor{wordcolor2}{rgb}{0.369,0.557,0.765}
\definecolor{wordcolor3}{rgb}{0.929,0.180,0.145}
\definecolor{wordcolor4}{rgb}{0.227,0.643,0.404}
\definecolor{wordcolor5}{rgb}{1, 0, 1}

\usepackage{lastpage}
\jmlrheading{23}{2024}{1-\pageref{LastPage}}{1/21; Revised 5/22}{9/22}{21-0000}{Dairui Liu}

\ShortHeadings{LIU, DU, Dong, Dong}{Transformers4NewsRec: A Transformer-based News Recommendation Framework}
\firstpageno{1}

\begin{document}

\title{Transformers4NewsRec: A Transformer-based News Recommendation Framework}

\author[1]{\textbf{Dairui Liu}}
\author[1]{\textbf{Honghui Du}}
\author[2]{\textbf{Boming Yang}}
\author[1]{\textbf{Neil Hurley}}
\author[1]{\textbf{Aonghus Lawlor}}
\author[2]{\textbf{Irene Li}}
\author[1]{\textbf{Derek Greene}}
\author[1]{\textbf{Ruihai Dong}}
\affil[1]{University College Dublin, Dublin, Ireland}
\affil[2]{The University of Tokyo, Tokyo, Japan}

\maketitle

\begin{abstract}
Pre-trained transformer models have shown great promise in various natural language processing tasks, including personalized news recommendations. To harness the power of these models, we introduce \textbf{Transformers4NewsRec}, a new Python framework built on the \textbf{Transformers} library. This framework is designed to unify and compare the performance of various news recommendation models, including deep neural networks and graph-based models. \textbf{Transformers4NewsRec} offers flexibility in terms of model selection, data preprocessing, and evaluation, allowing both quantitative and qualitative analysis.
\end{abstract}

\begin{keywords}
  Transformers, News Recommendations, Framework
\end{keywords}

\section{Introduction}

Personalized news recommendation has emerged as a crucial tool in delivering tailored content to users based on their preferences and reading behaviors. Hybrid models combining deep neural networks with powerful pre-trained language models (PLMs) have demonstrated significant improvements in various natural language processing tasks. However, applying these advances to news recommendation systems requires an adaptable framework that can handle the complexities of different data formats, user behaviors, and news content. This paper introduces \textbf{Transformers4NewsRec}, a flexible and scalable framework designed to leverage transformer-based architectures, traditional DL techniques, and graph-based methods to improve news recommendations.

\textbf{Transformers4NewsRec} is designed as a modular pipeline that allows researchers and developers to easily experiment with various models, datasets, and optimization strategies. The framework is built on powerful libraries, such as Transformers~\citep{wolf-etal-2020-transformers}, ensuring it can accommodate diverse experimental settings. 
Unlike existing frameworks that rely on zero-padding, we propose a novel concatenation-based batching method that significantly reduces redundant computations, leading to over 100\% improvements in training and evaluation speeds without sacrificing performance.
\textbf{Transformers4NewsRec} allows for easy experimentation with different models and datasets, providing a unified platform for performance comparison and generating topic-centric explanations to improve transparency.

\section{Framework: Transformers4NewsRec}

\begin{figure}[!t]
    \centering
    \includegraphics[width=0.8\linewidth]{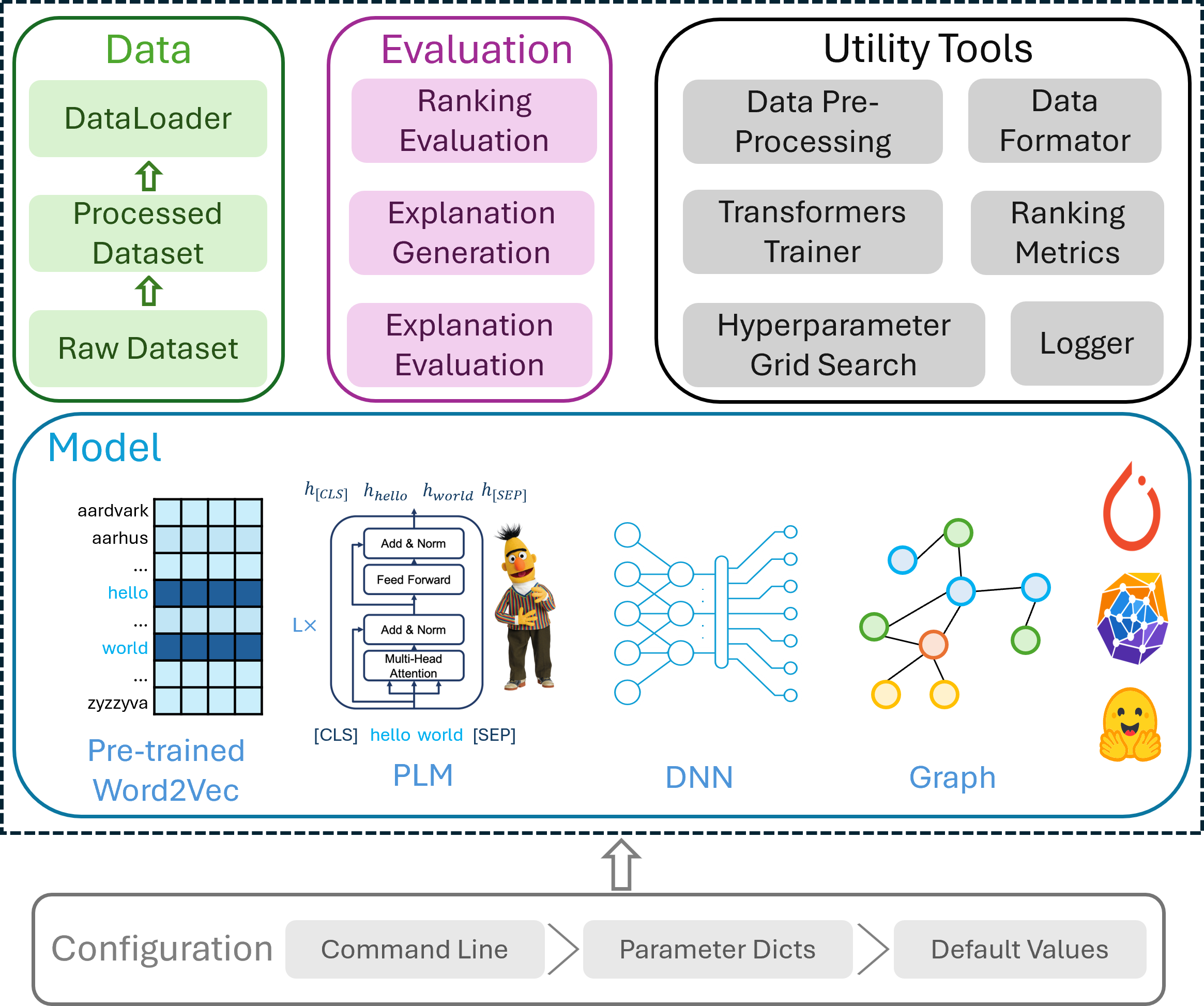}
    \caption{The overall architecture of the Transformers4NewsRec framework, illustrating its modular design. The framework integrates key components for data processing, model building, evaluation, and utility tools, enabling flexible experimentation and customization for news recommendation tasks.}
    \label{fig:transformers4newsrec_framework}
\end{figure}

The proposed framework for transformer-based news recommendation, as depicted in Figure~\ref{fig:transformers4newsrec_framework}, consists of four key components: \textbf{Data}, \textbf{Model}, \textbf{Evaluation}, and \textbf{Utility Tools}, offering a comprehensive solution for personalized news delivery.

The framework's configuration is highly flexible, utilizing a command-line interface that allows users to define model parameters, select models, and adjust experimental settings through command-line arguments, parameter dictionaries, or default values. This setup enables easy customization, allowing researchers and developers to test various models and datasets, optimizing performance efficiently.

\subsection{Data Module}

\begin{figure}[!t]
    \centering
    \includegraphics[width=\linewidth]{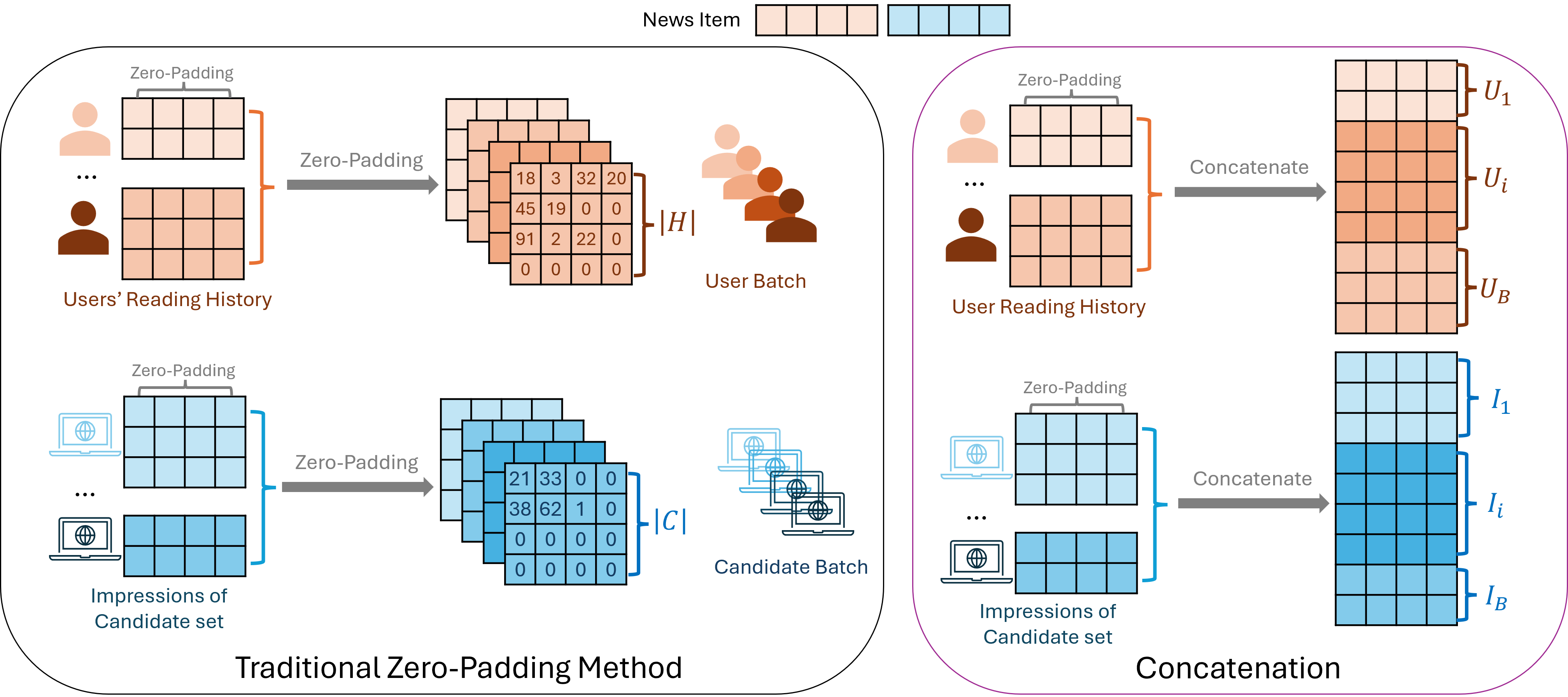}
    \caption{Comparison between zero-padding and concatenation methods for data batching. The left side illustrates the traditional zero-padding method, while the right side shows the proposed concatenation method, which improves efficiency by reducing redundant padding during batch creation, ultimately speeding up training and evaluation.}
    \label{fig:transformers4newsrec_concat}
\end{figure}

The \textbf{Data Module} initiates the pipeline and involves the transformation of raw inputs into processed datasets ready for model training and evaluation. The following components ensure streamlined data handling:
\begin{itemize}
    \item \textbf{Raw Dataset:} This includes the original news articles, user interaction histories, and any auxiliary contextual information. The raw data consists of features such as news titles, article bodies, categories, and user behavior data (e.g., impressions and clicks).

    \item \textbf{Processed Dataset:} After gathering the raw data, it is transformed into a structured format suitable for modeling. Preprocessing involves tokenization of news articles, feature extraction such as entity recognition and categorization, and preparation of graph data for models like GLORY. The \textit{FeatureMapper} class compiles these features into a feature matrix used during training and evaluation.

    \item \textbf{DataLoader:} This component feeds the processed data into the model, ensuring efficient batching and shuffling during both training and evaluation. Traditionally, batching involves zero-padding (as shown on the left side of Figure~\ref{fig:transformers4newsrec_concat}). However, we propose a more efficient concatenation method (right side of Figure~\ref{fig:transformers4newsrec_concat}), which minimizes redundant padding and accelerates both training and evaluation, particularly when using models with separate news and user encoders.
\end{itemize}

\subsection{Model Module}

The \textbf{Model Module} forms the core of the framework, integrating various modeling techniques to process news content and user interactions. This module supports flexible and scalable experimentation, leveraging popular libraries like PyTorch~\citep{pyTorch}, PyG~\citep{pyG}, and the Transformers~\citep{wolf-etal-2020-transformers} to accommodate different model configurations. News encoding, a critical task in personalized news recommendations, is handled using both traditional NLP techniques and state-of-the-art Transformer-based models.

For traditional word representations, the framework supports pre-trained word embedding methods, such as GloVe~\citep{pennington-etal-2014-glove}, which encode individual words in news articles as dense vectors. These embeddings form the foundation for more advanced models, providing a basic yet effective semantic understanding of the text.
More sophisticated approaches utilize Pre-trained Language Models (PLMs), such as BERT, which apply the self-attention mechanism to capture complex semantic relationships and contextual dependencies in the news content. This enables the model to form richer representations of news articles and user preferences, facilitating more accurate and contextually relevant recommendations.

Additionally, the Model Module integrates various Deep Neural Networks (DNNs), such as Convolutional Neural Networks (CNNs) and Gated Recurrent Units (GRUs). These networks are adept at learning intricate patterns in news content and user behavior, offering flexibility in designing custom models tailored to specific recommendation tasks.
The framework also incorporates graph-based models, which are essential for capturing relational information between users, news articles, and entities mentioned in the content. This graph-based approach enhances the model's ability to understand user preferences from a broader perspective, considering not just content-based interactions but also complex relationships within the user-news interaction network, making the recommendations more transparent and insightful.

\subsection{Evaluation and Utility Tools Module}

This section outlines two key framework components: (i) model evaluation after training and the tools that assist with data processing, training, evaluation; (ii) hyperparameter optimization. These components ensure effective model assessment and smooth system operation, from data preparation to result logging.

The \textbf{Evaluation Module} assesses the recommendation model's performance by measuring how well it ranks relevant news articles for users. Standard metrics such as AUC, MRR, nDCG@5, and nDCG@10 quantify ranking performance. In addition, the module generates and evaluates explanations, improving the transparency and interpretability of the recommendations.

The \textbf{Utility Tools Module} streamlines the entire pipeline, from data preprocessing to model training and optimization. It includes tools for transforming raw datasets into structured forms, handling tokenization, feature extraction, and embedding generation. The \textit{Transformers Trainer} fine-tunes pre-trained models, such as BERT, to optimize performance on news recommendation tasks. The \textit{Hyperparameter Grid Search} tool, powered by Optuna~\citep{optuna}, automates hyperparameter tuning, ensuring the best model configuration with minimal manual intervention.
Additional utilities include \textit{Ranking Metrics} and the \textit{Logger}. Ranking metrics are used during both training and evaluation to track the model's performance consistently, while the Logger monitors the entire process, recording key events, evaluation results, and any errors, facilitating experimentation and debugging. 

\section{Evaluation of Transformers4NewsRec}

In this section, we evaluate \textbf{Transformers4NewsRec} through three experiments. First, we compare the performance of six models under different configurations for news representation to validate the effectiveness of the proposed framework. Second, we estimate the efficiency of our proposed concatenation-based batching method versus the traditional zero-padding approach. 
Lastly, we present an initial analysis of topic-centric explanation generation.
\begin{table*}[t]
    \caption{Evaluation performance of six models on different feature sets.}
    \centering
    \begin{tabular}{l c c c c c c c c}
    \toprule
    \multirow{2}{*}{\textbf{Method}} & \multicolumn{4}{c}{\textbf{MIND-SMALL}} & \multicolumn{4}{c}{\textbf{MIND-LARGE}} \\
    \cline{2-9}
        & AUC & MRR & N@5 & N@10 & AUC & MRR & N@5 & N@10 \\
    \hline
    GloVe/Title & & & & & & & & \\
    NPA & 64.91 & 30.61 & 33.69 & 39.98 & 67.50 & 32.81 & 36.20 & 42.44 \\
    LSTUR & 66.15 & 31.53 & 34.72 & 40.99 & 68.96 & 33.54 & 37.14 & 43.54 \\
    NRMS & 66.47 & 31.66 & 35.04 & 41.31 & 68.19 & 33.20 & 36.70 & 43.12 \\
    NAML & 67.44 & 31.70 & 35.34 & 41.41 & 67.79 & 32.95 & 36.31 & 42.64 \\
    BATM-NR & 67.95 & 32.91 & 36.25 & 42.61 & 69.13 & 33.67 & 37.32 & 43.78 \\
    GLORY & \textbf{68.16} & \textbf{33.07} & \textbf{36.60} & \textbf{42.79} & \textbf{69.54} & \textbf{34.03} & \textbf{37.92} & \textbf{44.19} \\
    \hline
    GloVe/Title+Body & & & & & & & & \\
    NPA & 66.78 & 31.83 & 35.11 & 41.32 & 68.94 & 33.59 & 37.40 & 43.54 \\
    LSTUR & 67.58 & 32.29 & 35.70 & 41.90 & 69.87 & 33.87 & 37.67 & 43.93 \\
    NRMS & 67.97 & 32.78 & 36.24 & 42.38 & 69.67 & 34.12 & 38.07 & 44.29 \\
    NAML & 68.17 & 32.29 & 36.02 & 42.16 & 69.29 & 33.21 & 36.79 & 43.17 \\
    BATM-NR & 68.66 & 33.71 & 37.22 & 43.32 & 70.20 & 34.55 & 38.13 & 44.45 \\
    GLORY & \textbf{69.07} & \textbf{33.95} & \textbf{37.46} & \textbf{43.54} & \textbf{70.65} & \textbf{34.79} & \textbf{38.43} & \textbf{44.88} \\
    \hline
    BERT/Title & & & & & & & & \\
    NPA & 67.33 & 32.08 & 35.35 & 41.73 & 69.50 & 33.83 & 37.47 & 43.88 \\
    LSTUR & 67.74 & 31.86 & 35.30 & 41.69 & 69.51 & 33.84 & 37.32 & 43.76 \\
    NRMS & 68.50 & 32.67 & 36.27 & 42.46 & 68.97 & 33.22 & 36.75 & 43.24 \\
    NAML & 67.95 & 32.57 & 35.73 & 42.23 & 69.52 & 33.84 & 37.45 & 43.87 \\
    BATM-NR & 69.04 & 33.41 & 37.07 & 43.21 & 70.32 & 34.68 & 38.54 & 44.90 \\
    GLORY & \textbf{69.12} & \textbf{33.80} & \textbf{37.48} & \textbf{43.54} & \textbf{71.05} & \textbf{34.81} & \textbf{38.71} & \textbf{45.04} \\
    \bottomrule
    \end{tabular}
    \label{tab:results_districts}
\end{table*}

\subsection{Model Performance Evaluation}

To evaluate the effectiveness of the framework, we mainly select the following models to compare:
\begin{itemize}
    \item \textbf{NPA}~\citep{NPA2019ChuhanWu}: This method utilizes a personalized attention network to learn news and user representation.
    \item \textbf{NAML}~\citep{NAML2019ChuhanWu}: This method proposes the multi-view attention network learn representation based on different news article features.
    \item \textbf{LSTUR}~\citep{LSTUR2019MingxiaoAn}: This method jointly models users’ long-term and short-term interests by a GRU network.
    \item \textbf{NRMS}~\citep{NRMS2019ChuhanWu}: This method uses the multi-head self-attention network to learn user and news representation.
    \item \textbf{BATM-NR}~\citep{Liu2024TopicNR}: This method utilizes a Bi-level Attentive Topic Model to capture latent topic representations from both news content and user behaviors for better news recommendations.
    \item \textbf{GLORY}~\citep{GLORY}: This method constructs a global news graph based on user-interaction data, enhancing news representation through graph neural networks.
\end{itemize}
To evaluate \textbf{Transformers4NewsRec}, we applied it to six models: four traditional news recommendation models (NPA, LSTUR, NRMS, and NAML) and our two proposed models (BATM-NR and GLORY). These models were tested using three different feature sets: (i) GloVe~\citep{pennington-etal-2014-glove} embeddings using only the news title, (ii) GloVe embeddings using both the title and body, and (iii) BERT\citep{devlin-etal-2019-bert} embeddings using only the title.

Table~\ref{tab:results_districts} summarizes the performance of these models on the MIND-SMALL and MIND-LARGE datasets, with respect to metrics such as AUC, MRR, nDCG@5 (N@5), and nDCG@10 (N@10). Across all settings, BATM-NR and GLORY consistently outperform the classical models, particularly when both the news title and body are used. GLORY, in particular, achieves the best scores in most cases, demonstrating its superior ability to capture intricate user-news interactions with graph-based representations.

The inclusion of the news body improves model performance significantly, as it provides more comprehensive information about the news articles, enhancing the quality of personalized recommendations. Additionally, models utilizing pre-trained language models like BERT generally outperform those using traditional GloVe embeddings, further highlighting the benefits of leveraging transformers-based models for richer text representation.

\subsection{Efficiency Comparison of Batching Methods}

Efficient data processing is crucial for large-scale recommendation systems, especially when handling deep learning models. In this experiment, to measure the efficiency of data processing, we define \textbf{Training Speed} ($S_{\text{train}}$) as the number of samples processed per second during model training, while \textbf{Evaluation Speed} ($S_{\text{eval}}$) denotes the processing speed during evaluation.

\begin{table*}[t]
    \caption{Efficiency and Performance Comparison of Batching Methods (all experiments were conducted on an A100 40GB GPU in the MIND-SMALL dataset).}
    \begin{adjustbox}{width={\textwidth},keepaspectratio}
    \centering
    \begin{tabular}{l c c c c c c}
    \toprule
    \textbf{Batching Method} & \textbf{$S_{\text{train}}$} & \textbf{$S_{\text{eval}}$} & \textbf{AUC} & \textbf{MRR} & \textbf{N@5} & \textbf{N@10} \\
    \midrule
    NRMS/ZeroPad & 769.9/s & 880.8/s & 0.6568 & 0.3116 & 0.3428 & 0.4066 \\
    NRMS/Concat & 1630.0/s & 2622.5/s & 0.6647 & 0.3166 & 0.3504 & 0.4131 \\
    \textbf{Improvement} & \textbf{+111.77\%} & \textbf{+197.72\%} & \textbf{+1.20\%} & \textbf{+1.60\%} & \textbf{+2.22\%} & \textbf{+1.60\%} \\
    \midrule
    BATM-NR/ZeroPad & 767.4/s & 790.8/s & 0.6736 & 0.3220 & 0.3558 & 0.4186 \\
    BATM-NR/Concat & 1554.9/s & 2402.8/s & 0.6795 & 0.3291 & 0.3625 & 0.4261 \\
    \textbf{Improvement} & \textbf{+102.63\%} & \textbf{+203.88\%} & \textbf{+0.88\%} & \textbf{+2.20\%} & \textbf{+1.88\%} & \textbf{+1.79\%} \\
    \bottomrule
    \end{tabular}
    \end{adjustbox}
    \label{tab:batch_comparison}
\end{table*}

Table~\ref{tab:batch_comparison} presents a comparison of two models, NRMS and BATM-NR, evaluated using both the zero-padding and concatenation-based batching methods, as introduced in Figure~\ref{fig:transformers4newsrec_concat}. The concatenation method not only improves training speed ($S_{\text{train}}$) and evaluation speed ($S_{\text{eval}}$) by over 100\% but also maintains comparable performance. The substantial gains in speed highlight the computational efficiency of the concatenation-based method, making it highly beneficial for large-scale news recommendation tasks.

\subsection{Topic-Centric Explanation Example}

Beyond model performance, \textbf{Transformers4NewsRec} can generate topic-centric explanations for recommendations through the proposed BATM-NR model. By highlighting relevant news topics that influence a user's reading behavior, these explanations make the recommendation process more transparent. 
\begin{table}[hb]
\centering
\caption{Sampled example of topic-centric explanation generated by the BATM-NR model. Key terms for each topic are highlighted in different colors.}
\begin{adjustbox}{width={\textwidth},keepaspectratio}
\begin{tabular}{ccm{12cm}} 
\hline
\textbf{\textit{News ID}} & \textbf{\textit{Category}} & \textbf{\textit{News Content (Highlighted Topics)}} \\  
\hline
\multicolumn{3}{c}{\textbf{Selected Browsing History of User $U65494$}} \\  
\hline
N36934 & Movies & \textcolor{wordcolor3}{Francis} Ford Coppola's Slam on \textcolor{wordcolor2}{Marvel Films} Fuels Debate Sparked by \textcolor{wordcolor3}{Martin Scorsese}. Disparaging remarks by \textcolor{wordcolor3}{Francis} Ford Coppola have further inflamed the debate sparked by \textcolor{wordcolor3}{Martin Scorsese} and his criticism of \textcolor{wordcolor2}{Marvel} and other \textcolor{wordcolor2}{comic} book \textcolor{wordcolor2}{films}. At a press conference in \textcolor{wordcolor3}{Lyon, France}, where he was being honored at the Lumiere \textcolor{wordcolor2}{festival}, the Godfather director said he fully agreed with \textcolor{wordcolor3}{Scorsese's} assessment and went even further. \\  
\hline
N64137 & Finance & \textcolor{wordcolor2}{Verizon} to give its \textcolor{wordcolor4}{wireless} customers a year of \textcolor{wordcolor2}{Disney}. \textcolor{wordcolor2}{Verizon} announced Tuesday that it is teaming up with \textcolor{wordcolor2}{Disney} to give new and existing 4G LTE and 5G \textcolor{wordcolor4}{wireless} customers a free year of the new \textcolor{wordcolor2}{Disney streaming} service. The exclusive promotion coincides with the \textcolor{wordcolor2}{Disney} launch on Nov. 12. New \textcolor{wordcolor2}{Verizon} FIOS and 5G Home \textcolor{wordcolor4}{Internet} customers will also qualify for a year of \textcolor{wordcolor2}{Disney}. \\  
\hline
\multicolumn{3}{c}{\textbf{Recommended News Article from the Candidate News Pool}} \\  
\hline
N51180 & Movies & 13 Reasons Why Christian Navarro Slams \textcolor{wordcolor2}{Disney} for Casting the White \textcolor{wordcolor5}{Guy} in The Little \textcolor{wordcolor2}{Mermaid's}. \textcolor{wordcolor2}{Disney} has found their \textcolor{wordcolor5}{prince} and Navarro is not happy about it. Shortly after the announcement that someone had landed the role in the upcoming \textcolor{wordcolor5}{live} action retelling of \textit{The Little \textcolor{wordcolor2}{Mermaid}}, Navarro, 28, shared his dissatisfaction on social \textcolor{wordcolor2}{media}. \\  
\hline
\textbf{Explanation} & \multicolumn{2}{m{14.5cm}}{\textit{This recommendation is based on the user's browsing history, which indicates a strong interest in \textcolor{wordcolor3}{directors} like \textcolor{wordcolor3}{Francis Ford Coppola} and \textcolor{wordcolor3}{Martin Scorsese}, as well as \textcolor{wordcolor2}{films} like \textcolor{wordcolor2}{Marvel Films}. The article's focus on \textcolor{wordcolor2}{Disney} casting decisions and film industry controversies matches the user's previous reading habits.}} \\  
\hline
\end{tabular}
\end{adjustbox}
\label{table:history_and_candidate_news_example}
\end{table}

The framework analyzes user interaction history and identifies important topics that align with their preferences. The example in Table~\ref{table:history_and_candidate_news_example} demonstrates how a user's browsing history on specific topics related to movies and ``Disney'' informs the recommendation of related news articles. This process emphasizes transparency, providing users with a clear rationale behind each recommendation.

\section{Conclusion}
In this paper, we introduced \textbf{Transformers4NewsRec}, a flexible Python framework built on the \textbf{Transformers} library to improve personalized news recommendations. The framework integrates pre-trained transformer models, traditional deep neural networks, and graph-based models, offering a unified platform for both performance comparison and explanation generation. By incorporating efficient data processing techniques--the concatenation-based batching method, the framework enhances training and evaluation efficiency while maintaining robust recommendation performance. The framework's modular design enables seamless experimentation with different models and datasets, making it a valuable tool for researchers and developers. Through its ability to generate topic-centric explanations, the framework also provides transparency and interpretability for news recommendations.

\bibliography{reference}

\begin{thebibliography}{12}
\providecommand{\natexlab}[1]{#1}
\providecommand{\url}[1]{\texttt{#1}}
\expandafter\ifx\csname urlstyle\endcsname\relax
  \providecommand{\doi}[1]{doi: #1}\else
  \providecommand{\doi}{doi: \begingroup \urlstyle{rm}\Url}\fi

\bibitem[Akiba et~al.(2019)Akiba, Sano, Yanase, Ohta, and Koyama]{optuna}
T.~Akiba, S.~Sano, T.~Yanase, T.~Ohta, and M.~Koyama.
\newblock {O}ptuna: A next-generation hyperparameter optimization framework.
\newblock In \emph{The 25th ACM SIGKDD International Conference on Knowledge Discovery \& Data Mining}, pages 2623--2631, 2019.

\bibitem[An et~al.(2019)An, Wu, Wu, Zhang, Liu, and Xie]{LSTUR2019MingxiaoAn}
M.~An, F.~Wu, C.~Wu, K.~Zhang, Z.~Liu, and X.~Xie.
\newblock Neural news recommendation with long- and short-term user representations.
\newblock In \emph{Proceedings of the 57th Annual Meeting of the Association for Computational Linguistics}, pages 336--345. Association for Computational Linguistics, 2019.

\bibitem[Ansel et~al.(2024)Ansel, Yang, He, Gimelshein, Jain, Voznesensky, Bao, Bell, Berard, Burovski, Chauhan, Chourdia, Constable, Desmaison, DeVito, Ellison, Feng, Gong, Gschwind, Hirsh, Huang, Kalambarkar, Kirsch, Lazos, Lezcano, Liang, Liang, Lu, Luk, Maher, Pan, Puhrsch, Reso, Saroufim, Siraichi, Suk, Zhang, Suo, Tillet, Zhao, Wang, Zhou, Zou, Wang, Mathews, Wen, Chanan, Wu, and Chintala]{pyTorch}
J.~Ansel, E.~Yang, H.~He, N.~Gimelshein, A.~Jain, M.~Voznesensky, B.~Bao, P.~Bell, D.~Berard, E.~Burovski, G.~Chauhan, A.~Chourdia, W.~Constable, A.~Desmaison, Z.~DeVito, E.~Ellison, W.~Feng, J.~Gong, M.~Gschwind, B.~Hirsh, S.~Huang, K.~Kalambarkar, L.~Kirsch, M.~Lazos, M.~Lezcano, Y.~Liang, J.~Liang, Y.~Lu, C.~K. Luk, B.~Maher, Y.~Pan, C.~Puhrsch, M.~Reso, M.~Saroufim, M.~Y. Siraichi, H.~Suk, S.~Zhang, M.~Suo, P.~Tillet, X.~Zhao, E.~Wang, K.~Zhou, R.~Zou, X.~Wang, A.~Mathews, W.~Wen, G.~Chanan, P.~Wu, and S.~Chintala.
\newblock Pytorch 2: Faster machine learning through dynamic python bytecode transformation and graph compilation.
\newblock In \emph{Proceedings of the 29th ACM International Conference on Architectural Support for Programming Languages and Operating Systems, Volume 2}, ASPLOS '24, page 929–947, New York, NY, USA, 2024. Association for Computing Machinery.
\newblock ISBN 9798400703850.
\newblock \doi{10.1145/3620665.3640366}.
\newblock URL \url{https://doi.org/10.1145/3620665.3640366}.

\bibitem[Devlin et~al.(2019)Devlin, Chang, Lee, and Toutanova]{devlin-etal-2019-bert}
J.~Devlin, M.-W. Chang, K.~Lee, and K.~Toutanova.
\newblock {BERT}: Pre-training of deep bidirectional transformers for language understanding.
\newblock In \emph{Proceedings of the 2019 Conference of the North {A}merican Chapter of the Association for Computational Linguistics: Human Language Technologies, Volume 1 (Long and Short Papers)}, pages 4171--4186, Minneapolis, Minnesota, June 2019. Association for Computational Linguistics.
\newblock \doi{10.18653/v1/N19-1423}.
\newblock URL \url{https://aclanthology.org/N19-1423}.

\bibitem[Fey and Lenssen(2019)]{pyG}
M.~Fey and J.~E. Lenssen.
\newblock Fast graph representation learning with {PyTorch Geometric}.
\newblock In \emph{ICLR Workshop on Representation Learning on Graphs and Manifolds}, 2019.

\bibitem[Liu et~al.(2024)Liu, Greene, Li, Jiang, and Dong]{Liu2024TopicNR}
D.~Liu, D.~Greene, I.~Li, X.~Jiang, and R.~Dong.
\newblock Topic-centric explanations for news recommendation.
\newblock \emph{ACM Trans. Recomm. Syst.}, jul 2024.
\newblock \doi{10.1145/3680295}.
\newblock URL \url{https://doi.org/10.1145/3680295}.

\bibitem[Pennington et~al.(2014)Pennington, Socher, and Manning]{pennington-etal-2014-glove}
J.~Pennington, R.~Socher, and C.~Manning.
\newblock {G}lo{V}e: Global vectors for word representation.
\newblock In \emph{Proceedings of the 2014 Conference on Empirical Methods in Natural Language Processing ({EMNLP})}, pages 1532--1543, Doha, Qatar, Oct. 2014. Association for Computational Linguistics.
\newblock \doi{10.3115/v1/D14-1162}.
\newblock URL \url{https://aclanthology.org/D14-1162}.

\bibitem[Wolf et~al.(2020)Wolf, Debut, Sanh, Chaumond, Delangue, Moi, Cistac, Rault, Louf, Funtowicz, Davison, Shleifer, von Platen, Ma, Jernite, Plu, Xu, Le~Scao, Gugger, Drame, Lhoest, and Rush]{wolf-etal-2020-transformers}
T.~Wolf, L.~Debut, V.~Sanh, J.~Chaumond, C.~Delangue, A.~Moi, P.~Cistac, T.~Rault, R.~Louf, M.~Funtowicz, J.~Davison, S.~Shleifer, P.~von Platen, C.~Ma, Y.~Jernite, J.~Plu, C.~Xu, T.~Le~Scao, S.~Gugger, M.~Drame, Q.~Lhoest, and A.~Rush.
\newblock Transformers: State-of-the-art natural language processing.
\newblock In \emph{Proceedings of the 2020 Conference on Empirical Methods in Natural Language Processing: System Demonstrations}, pages 38--45, Online, Oct. 2020. Association for Computational Linguistics.
\newblock \doi{10.18653/v1/2020.emnlp-demos.6}.
\newblock URL \url{https://aclanthology.org/2020.emnlp-demos.6}.

\bibitem[Wu et~al.(2019{\natexlab{a}})Wu, Wu, An, Huang, Huang, and Xie]{NAML2019ChuhanWu}
C.~Wu, F.~Wu, M.~An, J.~Huang, Y.~Huang, and X.~Xie.
\newblock Neural news recommendation with attentive multi-view learning.
\newblock In \emph{International Joint Conference on Artificial Intelligence (IJCAI 2019)}, 2019{\natexlab{a}}.

\bibitem[Wu et~al.(2019{\natexlab{b}})Wu, Wu, An, Huang, Huang, and Xie]{NPA2019ChuhanWu}
C.~Wu, F.~Wu, M.~An, J.~Huang, Y.~Huang, and X.~Xie.
\newblock Npa: neural news recommendation with personalized attention.
\newblock In \emph{Proceedings of the 25th ACM SIGKDD international conference on knowledge discovery \& data mining}, pages 2576--2584, 2019{\natexlab{b}}.

\bibitem[Wu et~al.(2019{\natexlab{c}})Wu, Wu, Ge, Qi, Huang, and Xie]{NRMS2019ChuhanWu}
C.~Wu, F.~Wu, S.~Ge, T.~Qi, Y.~Huang, and X.~Xie.
\newblock Neural news recommendation with multi-head self-attention.
\newblock In \emph{Proceedings of the 2019 Conference on Empirical Methods in Natural Language Processing and the 9th International Joint Conference on Natural Language Processing (EMNLP-IJCNLP)}, pages 6389--6394, 2019{\natexlab{c}}.

\bibitem[Yang et~al.(2023)Yang, Liu, Suzumura, Dong, and Li]{GLORY}
B.~Yang, D.~Liu, T.~Suzumura, R.~Dong, and I.~Li.
\newblock Going beyond local: Global graph-enhanced personalized news recommendations.
\newblock \emph{Proceedings of the 17th ACM Conference on Recommender Systems}, 2023.

\end{thebibliography}

\end{document}